\documentclass[useAMS,usenatbib]{mn2e}
\usepackage{graphicx}
\usepackage{epstopdf}
\title[On the variability of quasars: a link between Eddington ratio and optical variability?]{On the variability of quasars: A link between Eddington ratio and optical variability?}
\author[B. C. Wilhite et al.]{Brian C. Wilhite$^{1,2}$\thanks{E-mail:wilhite@astro.uiuc.edu (BCW)},
Robert J. Brunner$^{1,2}$, 
Catherine J. Grier$^{1}$, Donald P.
\newauthor
Schneider$^{3}$, and
Daniel E. Vanden Berk$^{3}$\\
$^{1}$The University of Illinois, Department of Astronomy,
  1002 W. Green St., Urbana, IL 61801 USA\\
$^{2}$National Center for Supercomputing Applications, 1205 W. Clark St., Urbana, IL 61801 USA\\
$^{3}$The Pennsylvania State University, Department of Astronomy and Astrophysics, 525 Davey Lab, University Park, PA 16802}

\begin{document}


\pagerange{\pageref{firstpage}--\pageref{lastpage}} \pubyear{2006}

\maketitle

\label{firstpage}

\begin{abstract}

Repeat scans by the Sloan Digital Sky Survey (SDSS) of a 278-deg$^{2}$ stripe along the Celestial equator have yielded an average of over 10 observations each for nearly 8,000 spectroscopically confirmed quasars.  Over 2500 of these quasars are in the redshift range such that the C~{\sc iv} $\lambda$ 1549 emission line is visible in the SDSS spectrum.  Utilising the width of these C~{\sc iv} lines and the luminosity of the nearby continuum, we estimate black hole masses for these objects.  In an effort to isolate the effects of black hole mass and luminosity on the photometric variability of our dataset, we create several subsamples by binning in these two physical parameters.  By comparing the ensemble structure functions of the quasars in these bins, we are able to reproduce the well-known anticorrelation between luminosity and variability, now showing that this anticorrelation is independent of the black hole mass.  In addition, we find a correlation between variability and the mass of the central black hole.  By combining these two relations, we identify the Eddington ratio as a possible driver of quasar variability, most likely due to differences in accretion efficiency.

\end{abstract}


%
%

\section{Introduction\label{intro}}

The luminosities of quasars and other active galactic nuclei
(AGN) have been observed to vary on time-scales from hours to decades, and from X-ray to radio wavelengths.
The majority of quasars exhibit continuum variability on the order of 20 per cent on time-scales of months to years \citep[e.g.,][]{hook94,vandenberk04}.  In fact, variability has long been used as a selection criterion in creating quasar samples from photometric data \citep[e.g.,][]{ koo86, ivezic04, rengstorf04}.
Many simple correlations between photometric variability and various physical parameters have been known for decades.  These relationships are summarised by \citet{helfand01} and \citet{giveon99}.  Numerous studies \citep[e.g.,][]{hawkins02,devries03} have shown variability to correlate with time lag.  Anti-correlations have been found between variability and luminosity \citep[e.g.,][]{uomoto76,cristiani97}
and wavelength \citep[e.g.,][] {giveon99, trevese01}.   \citet[hereafter VB04]{vandenberk04}, using a sample of $\sim25,000$ quasars, confirmed these known correlations, and parametrized relationships between variability and time lag, luminosity, rest-frame wavelength and redshift.

Our understanding of the physics of the central black hole in quasars and active galactic nuclei (AGN) has long been tied to the variability of the quasar's luminosity. Intra-day variability in X-ray and optical light \citep[see, e.g.,][]{terrell67, kinman68, boller97} 
point towards a compact object, specifically a supermassive black hole, at the centre of an AGN.  More recently, reverberation mapping techniques have been used to determine the radius of the broad line region and, indirectly, to measure the mass of the central black hole \citep{blandford82, peterson93,kaspi00}.

Recently, \citet{wold07}, presented evidence suggesting that the photometric variability of quasars is linked to the mass of the central black hole.  Strength of variability for $\sim$ 100 quasars was approximated by finding the greatest single-epoch $R$-band deviation from the mean, by using optical light curves from the QUEST1 survey \citep{rengstorf04}.  Black hole masses were estimated from Sloan Digital Sky Survey \citep[SDSS;][]{york00} spectra using the 5100-\AA\ continuum luminosity and H$_{\beta}$ line width, as calibrated by \citet{vestergaard06}.  However, while a clear correlation between black hole mass and variability was found for time-scales greater than 100 days, \citet{wold07} were unable to reproduce the well-known inverse relationship between luminosity and variability for their sample as a whole.  This is likely due to the redshifting of the blue (and more variable) portion of the spectrum into the R-band at higher redshifts; this causes the high-luminosity quasars visible at higher redshifts to appear more variable than one might expect, based on previous published results, such as \citet{vandenberk04}.  With so few objects, it is difficult to truly isolate the dependence of  variability upon black hole mass, given the correlation of mass with luminosity, which is in turn tied to redshift and wavelength.   \citet{wold07} do report a correlation between black hole mass and variability at constant luminosity; these intriguing results might be more convincing, however, if the variability-luminosity dependence were in line with expectations, or if the sample were larger.  

With this in mind, we have examined the variability properties of a much larger sample of quasars from the Equatorial Stripe (see \S\,\ref{south}) of the SDSS.  With this sample, we are able to reproduce all of the well-established dependences of photometric variability, including the inverse correlation with luminosity, as well as the recently measured correlation with black hole mass.  We briefly describe the quasar sample in \S\,\ref{dataset} and the statistics used to measure variability in \S\,\ref{variability}.  In \S\,\ref{mbhlumvar}, we describe the methods used to estimate black hole masses, including the continuum- and line-fitting techniques, as well as the dependence of variability on black hole mass and luminosity.  Finally, we interpret our results in terms of physical models for quasars in \S\,\ref{discussion}, and we conclude in \S\,\ref{conclusions}.

Throughout the paper we assume the standard concordance cosmology 
with parameter values $\Omega_\Lambda = 0.7, \Omega_{M} = 0.3,$ and $H_{0}=70\ $km s$^{-1}$ Mpc$^{-1}$, consistent with \citet{spergel07}.

%
%

\section{The quasar dataset}\label{dataset}

\subsection{The Sloan Digital Sky Survey}\label{SDSS}

The Sloan Digital Sky Survey \citep{york00} is designed to image $\sim10,000\ $deg$^{2}$ and obtained follow-up spectra for roughly $10^{6}$ galaxies and $10^{5}$ quasars.  All imaging and spectroscopic observations are made with a dedicated 2.5-metre telescope \citep{gunn06} at the Apache Point Observatory in the Sacramento Mountains of New Mexico.  Imaging data are acquired by a 54-CCD drift-scan camera \citep{gunn98} equipped with the SDSS $u,g,r,i$ and $z$ filters \citep{fukugita96}; the data are processed by the PHOTO software pipeline \citep{lupton01}.   The photometric system is normalised such that SDSS magnitudes are on the AB system \citep{smith02, lupton99}.  A 0.5-metre telescope monitors site photometricity and extinction \citep{hogg01, tucker06}.   
Point source astrometry for the survey is accurate to less than $100$ milliarcseconds \citep{pier03}.  \citet{ivezic04} discuss imaging quality control.

Objects are targeted for follow-up spectroscopy as candidate galaxies \citep{strauss02,eisenstein01}, quasars \citep{richards02} or stars \citep{stoughton02}.  Targeted objects are grouped in 3-degree diameter `tiles' \citep{blanton03} and aluminum plates are drilled with 640 holes whose locations on the plate  correspond to the objects' sky locations.  Each plate is placed in the imaging plane of the telescope and plugged with optical fibres, which run from the telescope to twin spectrographs and are
assigned to roughly 500 galaxies, 50 quasars and 50 stars. 

SDSS spectra cover the observer-frame optical and near infrared, from 3900\AA--9100\AA\ at a spectral resolution of $\sim1900$.  Spectra are obtained in three or four consecutive 15-minute observations until an average minimum signal-to-noise ratio is met.  The spectra are calibrated by observations of 32 sky fibres, 8 reddening standard stars, and 8 spectrophotometric standard stars.  Spectra are flat-fielded and flux calibrated by the {\tt Spectro2d} pipeline.  Next, {\tt Spectro1d} identifies spectral features and classifies objects by spectral type \citep{stoughton02}.  Ninety-four percent of all SDSS quasars are identified spectroscopically by this automated calibration; the remaining quasars are identified through manual inspection.  Quasars are defined to be those extragalactic objects with broad emissions lines (full width at half maximum velocity width of $> 1000\ $km s$^{-1}$), regardless of luminosity.  \citet{vandenberk05} found that the SDSS targeting algorithm is 95 per cent complete to $i=19.1$, the limiting magnitude of the low-redshift survey.

\subsection{The SDSS equatorial stripe quasar dataset}\label{south}

During the autumn months, when the Northern Galactic Cap is unavailable for observation, the survey continually re-images a stripe centred on the celestial equator, as well as two `outrigger' stripes roughly 10 degrees North or South of the equator.  The equatorial scan, identified as Stripe 82, consists of a 2.5-degree wide stripe ranging from 309.2$^{\circ}$ to 59.8$^{\circ}$ Right Ascension, covering a total of 278 deg$^{2}$.  The SDSS Fifth Data Release \citep{adelman07} contains the 57 survey-quality imaging runs that cover Stripe 82, which were observed as part of regular SDSS-I operations through 2005 June.


We study only those objects observed spectroscopically, as they have been confirmed as quasars, and information about their black hole masses can be extracted directly from their spectra.  In this region, 7886 objects have been spectroscopically observed by the SDSS and confirmed to be quasars.  The majority of these quasars have been imaged by the SDSS between 8 and 12 times each, with an average of 9.5 (and a maximum of 27) observations per object.    

%
%

\section{Variability properties of equatorial stripe quasars}\label{variability}

\subsection{Construction of the structure function}\label{sf}

To measure the strength of the variability of our full sample and various subsamples, we use a standard formulation of the structure function \citep{diclemente96}:

\begin{equation}
V = \sqrt{(\frac{\pi}{2})\langle|\Delta{m}(\Delta{\tau})|\rangle^2-\langle\sigma_{n}^{2}\rangle},
\label{sfeq}
\end{equation}
where $\Delta{m}(\Delta{\tau})$ is the difference in magnitude between any two observations of a quasar, separated by $\Delta{\tau}$ in the quasars rest frame, and $\sigma_{n}^{2}$ is the square of the uncertainty in that difference (which is equal to the sum of the two individual observations' errors in quadrature).  The units of the structure function are magnitudes.  The means of these quantities are taken over 10 bins, ranging from 7 days to 700 days, of equal width in the logarithm of the time lag.

The structure function can be a useful tool, especially in comparing the relative variability of two subsamples of quasars, which is the primary approach employed in this paper.  However, the structure function is not an ideal measure in a statistical sense.  It assumes each point is statistically independent from all others, which is clearly not the case, as most quasars contribute more than one data point to each bin; this makes a true measurement of the error quite difficult.  We follow the lead of \citet{cristiani97} and \citet{rengstorf06} in estimating the error, by making the (known incorrect) assumption that the individual data points are independent, and ignoring covariance between points.  We then apply standard error propagation to Equation \ref{sfeq}, using the statistical error in the mean as the uncertainties for $\langle|\Delta{m(\Delta{\tau})}|\rangle$ and $\langle\sigma_{n}^{2}\rangle$ in each bin.  This leads to a slight overestimation of the uncertainty in the structure function \citep{rengstorf06}, which does not change any of our results; we defer a complete treatment of the covariance to a later paper.

\subsection{Structure function of the entire sample}\label{fullsample}

Figure \ref{Fig3.1} shows the structure function in all five SDSS photometric bands for the full sample of 7,886 quasars.  A comparison of these five structure functions shows that quasars are most variable in the $u$ band, and least variable in the $z$ band.  This is as one would expect, since it is well known that quasars vary more at blue wavelengths in the ultraviolet and optical \citep[see, e.g.,][]{wilhite05}.  To characterise these structure functions, we fit a power low to these data of the form:

\begin{equation}
V = \left(\frac{\Delta{\tau}}{\Delta{\tau}_{0}}\right)^{\alpha}.
\end{equation}
This is done by fitting a line to the logarithm of these data with the IDL function POLYFITW, weighting the points by the uncertainties calculated in \S\,\ref{sf}.  These values are translated into the corresponding values for a power law fit.
Parameters resulting from fits to the full-sample structure functions can be found in Table \ref{fullplfittab}.  

We employ a power law fit largely for historical reasons, as this has often been the functional form of choice \citep[e.g.][]{hook94,vandenberk04}.  While other functional forms may produce a better fit, in terms of the chi-squared parameter, we have found that the relative results in the values of  $V(\Delta{\tau}=100)$ are completely unchanged.  Thus, we use a power law to fit our structure functions, to allow for comparisons with previous work.

Using this functional form, the parameter $\Delta{\tau}_{0}$ should not be construed as a natural time-scale for variability; it is simply the time at which the structure function reaches a value of one magnitude.  For structure functions of roughly equal power law slope ($\alpha$), however, a lower value of $\Delta{\tau}_{0}$ corresponds to a greater strength of variability.  Thus, as seen in Table \ref{fullplfittab}, values for $\Delta{\tau}_{0}$ increase as one proceeds redward in wavelength, indicating declining variability, as seen in Figure \ref{Fig3.1}.  However, in cases where the power law slopes of multiple structure functions are not equal, neither $\Delta{\tau}_{0}$ nor $\alpha$ gives a reliable measure of the strength of variability.  This can be better achieved by evaluating the function at some chosen value of  $\Delta{\tau}$; we choose a value of 100 days, as this falls near the centre of the range of time lags sampled by our data.
We shall, therefore, use the $V(\Delta{\tau}=100)$ parameter as a proxy for the strength of variability when comparing the structure functions for samples constructed by binning in black hole mass and continuum luminosity.  The uncertainties quoted for values of $V(\Delta{\tau}=100)$ in Tables 1 and 3 are obtained by using the uncertainties in $\Delta{\tau}_{0}$ and $\alpha$ returned by IDL and employing standard error propagation.
 
\section{Relating variability to luminosity and black hole mass}\label{mbhlumvar}

\subsection{Estimating M$_{BH}$ and $\lambda L_{\lambda} (1450)$}\label{mbhlumcalc}

A total of 2,531 of our quasars are at a sufficiently high redshift ($z > 1.69$) such that the entire C~{\sc iv} line profile is covered by the SDSS spectra.  
For these 2,531 high-redshift quasars, the median redshift roughly 2.1 with an mean near 2.5.  Black hole masses can be estimated for these objects using the C~{\sc iv} line dispersion (i.e., line width) and the 1450-\AA\ continuum luminosity.

Masses may be estimated through the widths of other lines, usually Mg~{\sc ii} and H$_{\beta}$, but this requires more complicated template fitting, and extends the redshift range of objects under study, possibly complicating our results.  Thus, for this initial work, we limit ourselves here to estimates which utilise the C~{\sc iv} line.  

Single-epoch black hole mass estimation techniques rely on the assumption that the gas in the broad emitting line region is in virial equilibrium, and that the velocity of the gas ($v$) is a reflection of the mass of the black hole ($M_{BH}$), and the distance from the black hole to the emitting gas ($R$):

\begin{equation}
M_{BH} = f\frac{R (\Delta{v})^{2}}{G},
\end{equation}
where $f$ is a dimensionless factor of order unity that depends upon the precise geometry of the broad line region.  In this scenario, the width of a given emission line is related to the gravitational potential of the central source; thus the line width serves as a proxy for the gas's orbital velocity.  Though plausible other scenarios exist in which the line width is not dominated by gravity, but some other factor such as radiation pressure, we here assume that these line widths provide information relating to the mass of the central black hole.

Considerable work has been done recently to calibrate the radius-luminosity relationship, applying reverberation mapping techniques to a collection of nearby Seyfert galaxies \citep[e.g.,][]{peterson04,kaspi05,bentz06}.
Once a reliable calibration has been determined, a single-epoch measurement of the luminosity may be used to estimate the radius of the broad-line region.  Here, we follow the prescription first described by \citet{vestergaard02} and later refined by \citet{vestergaard06}:

\begin{eqnarray*}
{\rm log\ M_{BH}(C~\textsc{iv})} & = & {\rm log} \left [ \left ( \frac{{\rm \sigma(C~\textsc{iv})}}{{\rm 1000\ km\ s^{-1}}} \right )^{2}
\left ( \frac{{\rm \lambda{L_{\lambda}(1450\AA)}}}{{\rm 10^{44}\ erg\ s^{-1}}} \right )^{0.53} \right ]\\ 
 & & + (6.73 \pm 0.01).
\end{eqnarray*}

This particular estimate for M$_{\rm BH}$ employs the non-parametric dispersion ($\sigma$) of the C~{\sc iv} line.  To measure the dispersion, we use the techniques developed previously by \citet{wilhite06}.  A linear fit is applied to the local continuum, using only the portion of the spectrum corresponding to the rest-frame intervals 1472 \AA--1487 \AA and 1685 \AA--1700 \AA. This fit is subtracted from the full spectrum to isolate the flux in the C~{\sc iv} line.  The median of the line is calculated and the dispersion in the line is calculated around the median measurement of the line centre.  Uncertainties are estimated by a Monte Carlo technique which involves repeatedly adding Gaussian noise to the spectrum and re-measuring the dispersion.  A full description of this technique can be found in \citet{wilhite06}.

To measure the continuum luminosity at 1450 \AA, we simply take the mean flux in a 10-\AA\ region centred on 1450 \AA\ and use it to calculate the intrinsic luminosity in our assumed cosmology.  The uncertainty in $\lambda L_{\lambda} (1450\AA)$ is estimated by calculating the error in the mean flux for the region and using standard error propagation.  The median value for this uncertainty in $\lambda L_{\lambda} (1450\AA)$ is $2.2 \times 10^{44}$ erg s$^{1}$, indicating the uncertainties are at the roughly 5\% level.  The uncertainty in black hole mass is estimated by standard propagation of the uncertainties in  $\lambda L_{\lambda} (1450\AA)$ and $\sigma(C~\textsc{iv})$, which yields a median M$_{\rm BH}$ uncertainty of $9 \times 10^{7}$ M${\sun}$, at the 10\% to 15\% level.

It should be noted that these mass estimates suffer from large systematic and random uncertainties.  \citet{baskin05} demonstrated that black hole mass estimates involving C~\textsc{iv} line width may be biased, perhaps with systematic over or underestimates of mass by a factor of a few.  Additionally, \citet{kelly07} find that the distribution of single-epoch mass estimates is likely too broad, relative to the presumed intrinsic distribution, while \citet{shang07} suggest that outflows may play a significant role in broad line widths. \citet{vestergaard06} state that UV-based single-epoch mass estimates, based on comparisons with their reverberation-mapping counterparts, are good to within a factor of a few.  As we are binning our objects in black hole mass and comparing the variability amplitudes of these subsamples (see \S\,\ref{mbhlumbin}), rather than studying individual objects, our results should be robust against both random and systematic uncertainties, provided that the C~\textsc{iv} line width is related to the mass of the central black hole.


\subsection{Binning in black hole mass and luminosity}\label{mbhlumbin}

Figure \ref{Fig3.2} displays the distribution in continuum luminosity versus the estimated black hole mass for the majority of the 2,531 quasars with measured C~\textsc{iv} emission lines from the SDSS Equatorial Stripe.  For continuum luminosity, we simply use the value for $\lambda L_{\lambda} (1450\AA)$ determined in \S\ref{mbhlumbin}.  To investigate the dependence of variability on black hole mass, we subdivide the luminosity-black hole mass plane into the six bins as marked on the distribution of quasars in Figure \ref{Fig3.2}.  The median continuum luminosity, redshift and black hole mass for each bin are listed in Table \ref{bininfotab}.  Also listed are the boundaries in black hole mass and luminosity for each bin.  To keep quasars with unreasonably low estimates of black hole mass from affecting the results, we do not include any of the 226 quasars with $M_{BH} < 10^{6} M_{\sun}$.   The vast majority of these low estimates are due to broad absorption of either the continuum or the emission line itself, and are unlikely to be accurate estimates of the true black hole mass.  Additionally, no bins include those 322 quasars with estimated masses above $2 \times 10^{9} M_{\sun}$, as there are simply too few in any region of  $\lambda L_{\lambda} (1450\AA)$---$M_{BH}$ parameter space to allow for a reliable measurement of their ensemble variability.  The quasars with estimated black hole mass less than $10^{6} M_{\sun}$ or greater than $2 \times 10^{9} M_{\sun}$ are not shown in Figure \ref{Fig3.2}.

For each bin, we calculate the structure functions for all five bands of the quasars in that bin.  All thirty structure functions (six bins times 5 bands) are shown in Figure \ref{Fig3.3}.    Each structure function demonstrates the familiar relation between wavelength and variability; the $u$ band in each bin shows the largest amplitude in its structure function, while the $z$-band measurements show the least variability.  The structure functions shown in Figure \ref{Fig3.3} have only nine points in $\Delta{\tau}$, rather than the ten seen in Figure \ref{Fig3.1}; the high-redshift nature of these quasars (which is necessary to observe C~\textsc{iv}) results in the largest rest-frame time lag bin containing no observations, after one translates from the observed frame to the quasar's rest frame.

One quickly notices the large level of uncertainty in virtually all of these 30 structure functions in the fifth bin in $\Delta{\tau}$, which is at approximately 60 days.  This is due to the lack of observations separated by 180 days in the observed frame; this bin spans $180\ \mathrm{days}/(1 + \langle z\rangle)$, where $\langle z\rangle$ is the mean redshift at which C~\textsc{iv} is observable (i.e., $z \approx 2.5$).  Additionally, in certain time-lag bins, a reliable measurement of the variability cannot be made, as the average uncertainty is greater than the average variability.  This is seen most often in $u$- and $z$-band structure functions, as those bands have the lowest signal-to-noise flux determinations.

By comparing the structure functions of quasars from adjacent bins in Figure \ref{Fig3.2}, we can isolate the dependences of variability upon luminosity and black hole mass.  For example, the left-hand panel of Figure \ref{Fig3.4} shows the $g$-band structure functions for the quasars from bins 1, 2 and 3.  Bin 1 quasars are clearly more variable than those in bin 2, which are, in turn, more variable that those in bin 3.  Table 3 shows the results of the power-law fits to these structure functions (as well as those representing the quasars in bins 4, 5 and 6).  The progression from high to low variability, as one travels from bin 1 to bin 3, seen in Figure \ref{Fig3.2} is reflected in the values for $V(\Delta{\tau}=100)$ for those bins.  In the right-hand panel of Figure \ref{Fig3.2}, the same relation is observed for quasars at higher black hole mass.  Quasars in bin 4 are of lower luminosity than those in bin 5, and are also more variable.

These results are not surprising, in that an anticorrelation between luminosity and variability has been known for decades.  However, this shows, for the first time, that this dependence exists independent of black hole mass, a property known to be correlated with luminosity.

By comparing bins with quasars of similar luminosity, but different black hole mass, one can isolate the dependence of variability on black hole mass.  This is seen with bins 2 and 4, as they cover the same range in luminosity, but bin 2 contains objects with $M_{BH} < 5 \times 10^{8} {\rm M}_{\sun}$, while bin 4 contains quasars with  between $5 \times 10^{8} {\rm M}_{\sun} < M_{\rm BH} < 10^{9} {\rm M}_{\sun}$.  The left-hand panel of Figure \ref{Fig3.5} shows these two bins' $g$-band structure functions, which indicate that the objects in bin 4--or those with the higher average black hole masses--are more variable than those in bin 2.  This is also reflected in their respective values of $V(\Delta{\tau}=100)$ listed in Table 3.

This same trend can be seen by comparing the three highest-luminosity bins: 3, 5 and 6.  In the right-hand panel of Figure \ref{Fig3.5} and Table 3, it can be seen that variability appears to increase with increasing black hole mass.  The increase is especially clear when one compares bin 3 with bin 6, the highest-black-hole-mass bin in our sample.

%
%

\section{Discussion}\label{discussion}

By isolating the dependence of variability upon luminosity and black hole mass, we are, in effect, able to probe the dependence of variability upon the Eddington ratio, $L_{bol}/L_{Edd}$.  The Eddington ratio of a quasar is a comparison of the actual bolometric luminosity, $L_{bol}$, to the Eddington luminosity, L$_{\rm Edd}$, which is the maximum stable luminosity at which accretion can occur.  However, as we are measuring the optical luminosity, we can recast this as:

\begin{equation}
L_{opt} = \varepsilon L_{bol},
\label{optbolratio}
\end{equation}
where $\varepsilon$ represents the fraction of the bolometric luminosity emitted in the optical.  This is likely to be a function of the bolometric luminosity; however, recent measurements for quasars with $L_{bol} > 10^{10} L_{\sun}$  have shown this value to be approximately 0.1 \citep{hopkins07,richards06}.
Furthermore, since the Eddington luminosity is directly proportional to black hole mass \citep{rees84}, we have that $L_{bol}/L_{Edd} \sim L_{opt}/M_{BH}$.

Characteristic Eddington ratios have been calculated for each bin and are provided in Table \ref{bininfotab}.  These values do not represent an average $L_{bol}/L_{Edd}$ for the bin, but rather the Eddington ratio one obtains from the average values for $\lambda L_{\lambda} (1450\AA)$ and M$_{\rm BH}$ also given in Table \ref{bininfotab}.  The black hole mass is converted to an Eddington luminosity through the familiar $L_{Edd} = 1.3 \times 10 ^{38} (M/{\rm M)_{\sun}}$ erg s$^{-1}$.  To get the Bolometric luminosity, we use the $L_{bol} \sim 9\times \lambda L_{\lambda} (5100\AA)$ relation used in \citet{kaspi00} and \citet{kollmeier06} and combine it with the $\alpha_{\nu} = 0.44$ quasar spectral slope of \citet{vandenberk01} to get a new relation for the continuum near the C~\textsc{iv} line: $L_{bol} \sim 5\times\lambda L_{\lambda} (1450\AA)$.  Five of the six bins have $L_{bol}/L_{Edd}$ between 0.1 and 1, as did the vast majority of objects in \citet{kollmeier06}.  Even Bin 3, with a value of $L_{bol}/L_{Edd}$ greater than 1 is not unreasonable; a number of objects studied in \citet{kollmeier06} were calculated to have super-Eddington luminosities. At any rate, the Eddington ratios calculated in Table \ref{bininfotab} should primarily be used as a means for comparing the relative Eddington ratios of the quasars in different bins.

By combining the established (and herein reproduced) inverse dependence of variability upon optical luminosity with the newly demonstrated correlation of variability with black hole mass, we find that variability appears to be inversely related to the Eddington ratio.  Quasars with higher Eddington ratios are less variable than those with lower Eddington ratios.  This suggests that the well-known anticorrelation of variability with luminosity may in fact simply be a side effect of a primary anticorrelation between variability and the Eddington ratio.

In Figure \ref{Fig3.1}, lines of constant Eddington ratio are simply lines with intercept zero.  In this plane, a higher Eddington ratio corresponds to a line with smaller positive slope.  We have avoided binning objects by their Eddington ratio in this paper, simply because the shapes of those bins would not lend themselves to easy comparisons.  We would, however, point out that bin 3 is the bin with the highest mean Eddington ratio.  As seen in Table \ref{mbhlumplfittab}, the quasars in bin 3 are also seen to be the least variable, with the lowest value for $V(\Delta{\tau}=100)$.

To interpret our hypothesised relationship between optical variability and the Eddington ratio, we use the theoretical relationship between the luminosity of a quasar and its accretion rate:
\begin{equation}
L_{bol} = \eta \dot M c^{2},
\label{radeff}
\end{equation}
where $\eta$ is a measure of the radiative efficiency of the quasar and is dependent on the specific physical parameters used to model the black hole~\citep[see, e.g.,][for detailed calculations]{krolikbook}. The two canonical values correspond to the Schwarzschild black hole, which has $\eta \approx 0.06$, and the Kerr black hole, which has $\eta = 0.42$. Given our lack of knowledge about the physical parameters of the supermassive black holes that power quasars, the general practice is to adopt a value that lies between these two extremes, i.e., $\eta \sim 0.1$.  

By combining Equations \ref{optbolratio} and \ref{radeff}, we have the simple model in which the optical luminosity is related to the accretion rate ($\dot M$), the radiative efficiency ({$\eta$) and the fraction of the bolometric luminosity that is emitted in the optical ($\varepsilon$):
\begin{equation}
L_{opt} = \varepsilon \eta \dot M c^{2}
\label{simplemodel}
\end{equation}
In light of Equation~\ref{simplemodel}, changes in the optical luminosity of a quasar can be driven either by a change in $\epsilon$, $\eta$, or $\dot M$.  A varying value of $\varepsilon$ would require radical changes of a quasar's spectral shape across multiple wavelength regimes.  A varying $\eta$ would require the nature of an individual black hole to change with time.  On the rest-frame time-scales of our observations, it is unlikely that either of these two would be comparable to variations in the accretion flow, which should naturally occur due to the dynamics of the entire accretion process.  



If we assume that variations in the optical luminosity of the quasar are tied to variations in the accretion rate, this can be interpreted as a link between the optical variability of a quasar and its `age'.  In the cocoon model \citep[see, e.g.,][]{haas04, hopkins05}, quasars become observable in the optical at high accretion rate (after feedback `blows away' enshrouding gas and dust), and fade away when the accretion rate drops.  The Eddington ratio, therefore, could be construed as a proxy for the age of the quasar, or more precisely, the time since the quasar became observable in the optical portion of the spectrum.  \citet{martini03} describe one possible test for measuring quasar lifetimes in models such as this, employing large, multi-epoch surveys.

At constant black hole mass, optical luminosity could provide a measure of the gas that is available for accretion onto the black hole.  Therefore, we might expect that younger quasars are more luminous because they have a greater fuel supply.  Similarly, when comparing two quasars with the same optical luminosity, the quasar with the larger black hole mass would be older--its lower Eddington ratio is indicative of it having burned through much of its once-larger fuel supply.  Thus, when comparing populations of quasars (as in our bins in $L_{opt}$ and $M_{BH}$), the greater variability seen in the lower luminosity objects would be a consequence of a dwindling fuel supply.  As less gas is available, the rate at which the gas is supplied to the black hole varies more, much like the flickering of a dying fire.  Either way, the possibility that variability is tied to the Eddington ratio, which is in essence a measure of the efficiency of a quasar, is an intriguing one.

Both panels of Figure \ref{Fig3.5} appear to demonstrate that black hole mass is related to variability at larger time lags.  This is also seen in Table \ref{mbhlumplfittab}, which shows that the high black hole mass bins not only have smaller values of $V(\Delta{\tau}=100)$, but also larger power law slopes, indicating that the differences in variability will be more prominent at longer time lags.  This agrees with \citet{wold07}, who saw little correlation between variability and black hole mass for a sample of observations with time separations less than 100 days, but a clear correlation between the two for $\Delta{\tau}$ greater than 100 days.  This apparent increase in the effect of back hole mass on longer time-scale variability clearly indicates the need for longer observed time baselines.  The results presented herein only use data from the completed SDSS-I survey.  The ongoing SDSS-II will ultimately add another three years to this baseline, for an average increase in the maximum rest-frame $\Delta{\tau}$ of roughly one year for each quasar.  

The analysis in this paper focused on the C~\textsc{iv} sample, which consists only of quasars with $z > 1.69$, as C~\textsc{iv} is blueward of the SDSS spectral response at lower redshifts. The remaining, lower redshift quasars can be analysed in a similar manner, however, by utilising other emission lines, such as Mg~\textsc{ii} or H$_{\beta}$.  Not only would this analysis nearly triple the number of quasars studied, but it would also extend the redshift baseline of our sample, thereby allowing us to test the hypothesised relationship between optical variability and accretion rate at other cosmic epochs.


%
%

\section{Conclusions}\label{conclusions}

In this paper, we have studied the ensemble variability properties of almost 8,000 spectroscopically identified quasars from the Sloan Digital Sky Survey Equatorial Stripe.  These objects have been observed an average of over ten times each.  By using their  C~{\sc iv}  line dispersions and nearby continuum luminosities, we have estimated black hole masses for approximately 2,500 of these quasars.  We have binned these quasars in luminosity and black hole mass and examined the variability properties of the quasars in each bin.  We have been able to:

(1) Reproduce the well-known anticorrelation between luminosity and variability, and

(2) Detect a correlation between variability and black hole mass.  

By combining (1) and (2), it appears that variability is inversely related to the Eddington ratio in quasars.  This points to variability being related to the quasar's accretion efficiency.  Given that the relation with black hole mass is more evident at longer time lags, we believe future studies involving longer time baselines will shed more light on this new result.

B. C. W. and R. J. B. would like to acknowledge support from Microsoft Research, the University of Illinois, and NASA through grants NNG06GH156 and NB 2006-02049. The authors made extensive use of the storage and computing facilities at the National Center for Supercomputing Applications and thank the technical staff for their assistance in enabling this work.

Funding for the SDSS and SDSS-II has been provided by the Alfred P. Sloan Foundation, the Participating Institutions, the National Science Foundation, the U.S. Department of Energy, the National Aeronautics and Space Administration, the Japanese Monbukagakusho, the Max Planck Society, and the Higher Education Funding Council for England. The SDSS Web Site is {\tt http://www.sdss.org/}.

The SDSS is managed by the Astrophysical Research Consortium for the Participating Institutions. The Participating Institutions are the American Museum of Natural History, Astrophysical Institute Potsdam, University of Basel, University of Cambridge, Case Western Reserve University, University of Chicago, Drexel University, Fermilab, the Institute for Advanced Study, the Japan Participation Group, Johns Hopkins University, the Joint Institute for Nuclear Astrophysics, the Kavli Institute for Particle Astrophysics and Cosmology, the Korean Scientist Group, the Chinese Academy of Sciences (LAMOST), Los Alamos National Laboratory, the Max-Planck-Institute for Astronomy (MPIA), the Max-Planck-Institute for Astrophysics (MPA), New Mexico State University, Ohio State University, University of Pittsburgh, University of Portsmouth, Princeton University, the United States Naval Observatory, and the University of Washington.

%
%

\onecolumn
\clearpage

%
%



\begin{figure}
\includegraphics{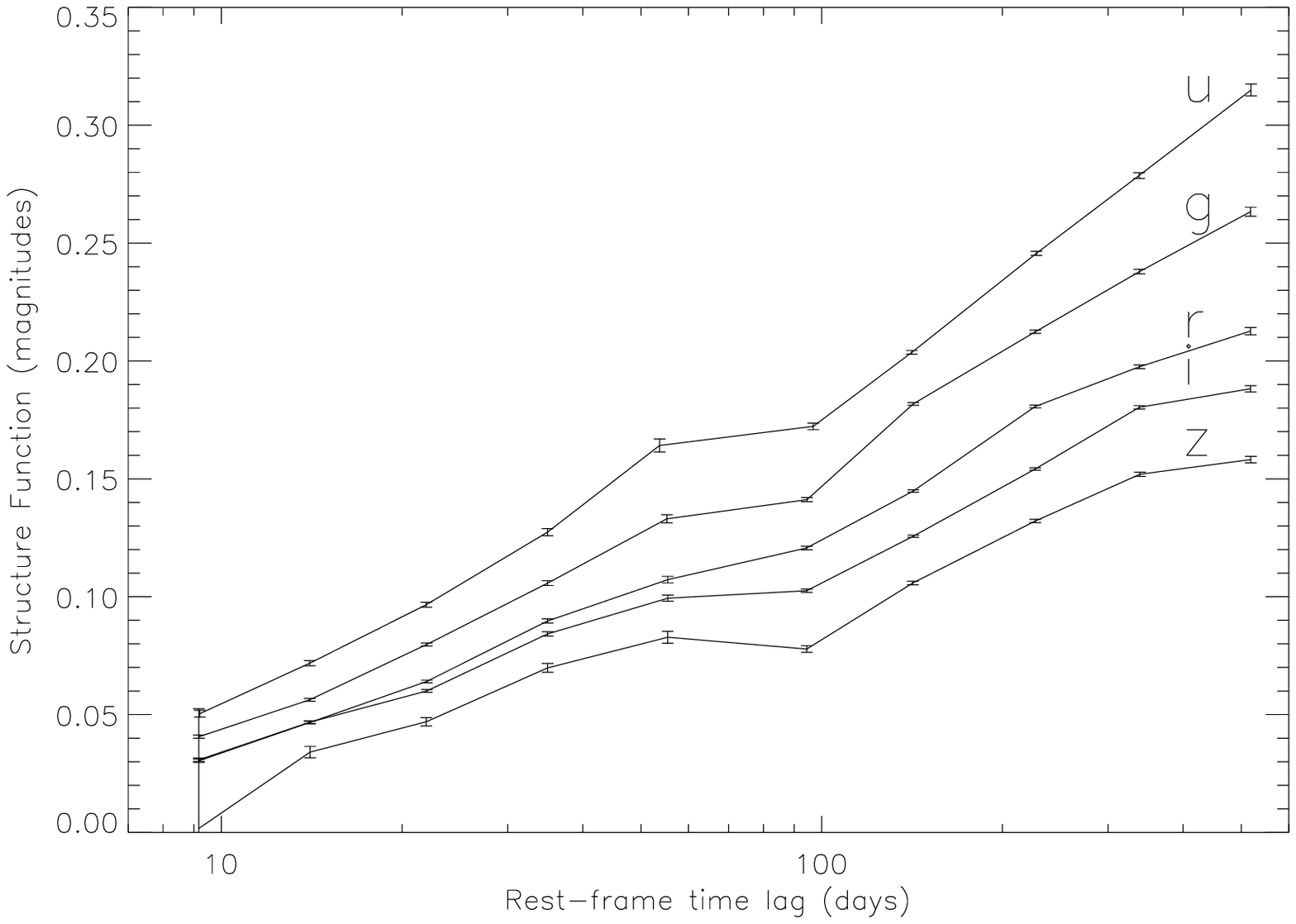}
\caption{Variability as a function of time (the Structure Function) for the full sample of 7,886 quasars in all five SDSS photometric bands.
\label{Fig3.1}}
\end{figure}
\clearpage

\begin{figure}
\includegraphics{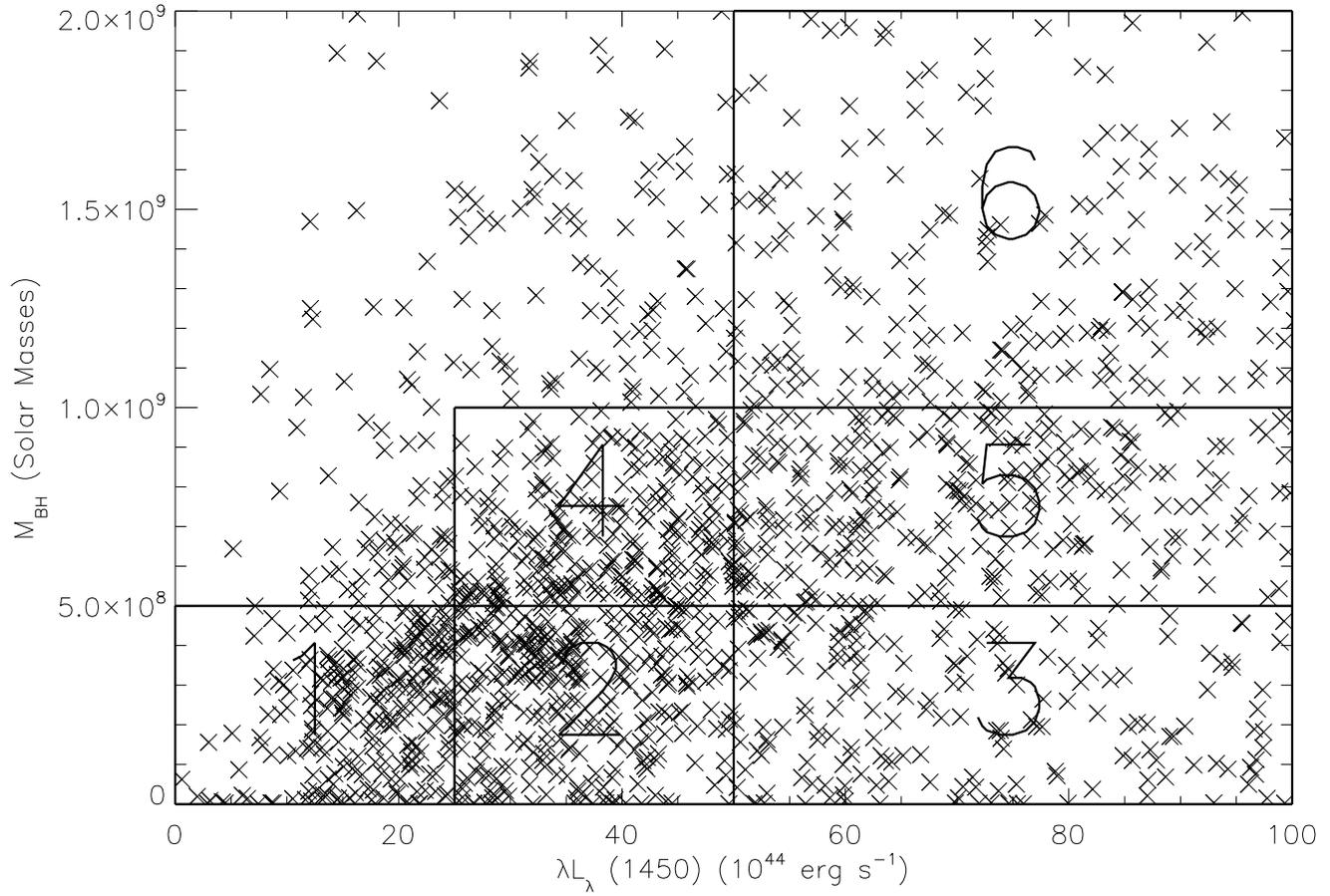}
\caption{Distribution in luminosity and black hole mass for the 2,531 quasars for which black hole masses have been estimated.  The lines overdrawn separate the objects into six bins, which are marked, used for the study of ensemble variability.
\label{Fig3.2}}
\end{figure}
\clearpage

\begin{figure}
\includegraphics{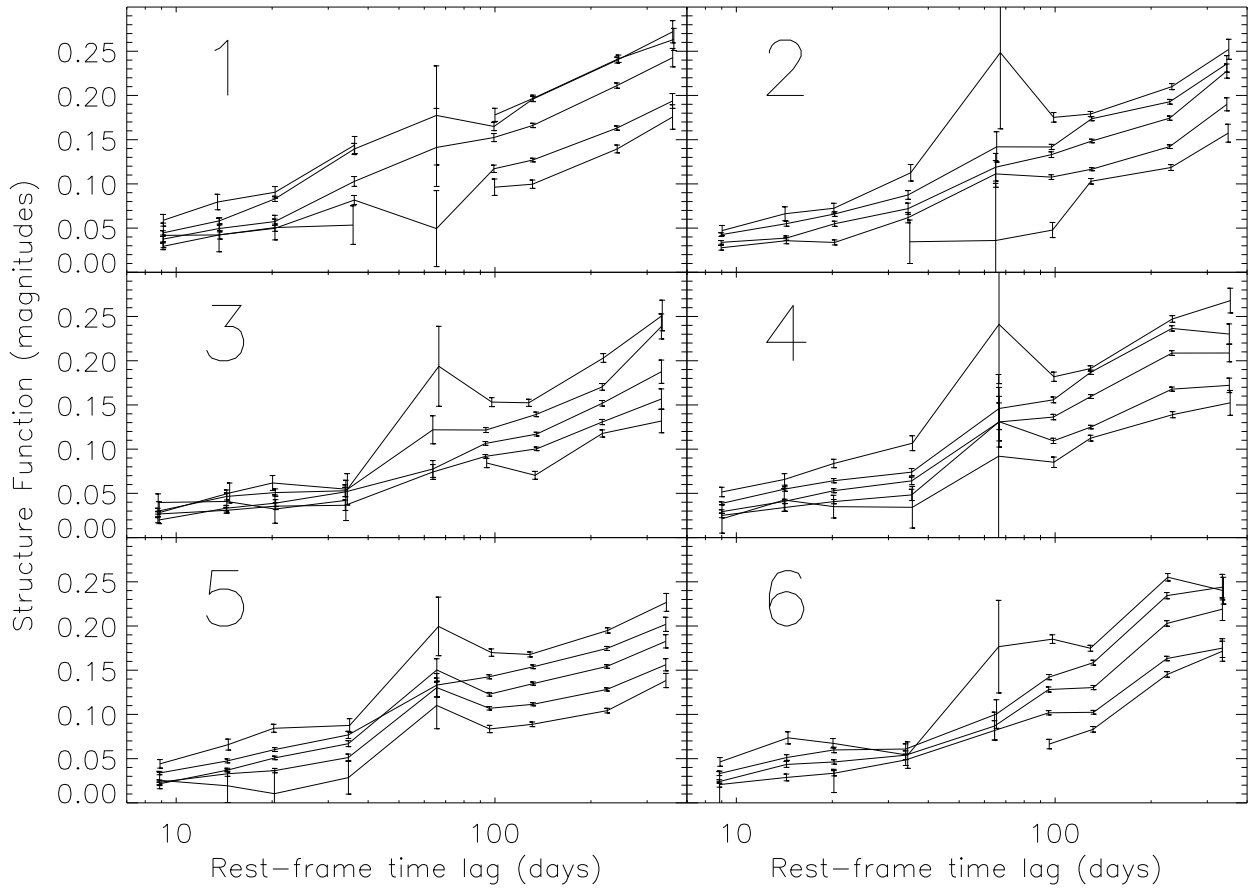}
\caption{Variability as a function of time (the Structure Function) in all five photometric bands and all six bins in luminosity and black hole mass.  In each panel, the curves represent, from most to least variable, the SDSS $u, g, r, i,$ and $z$ bands.
\label{Fig3.3}}
\end{figure}
\clearpage

\begin{figure}
\includegraphics{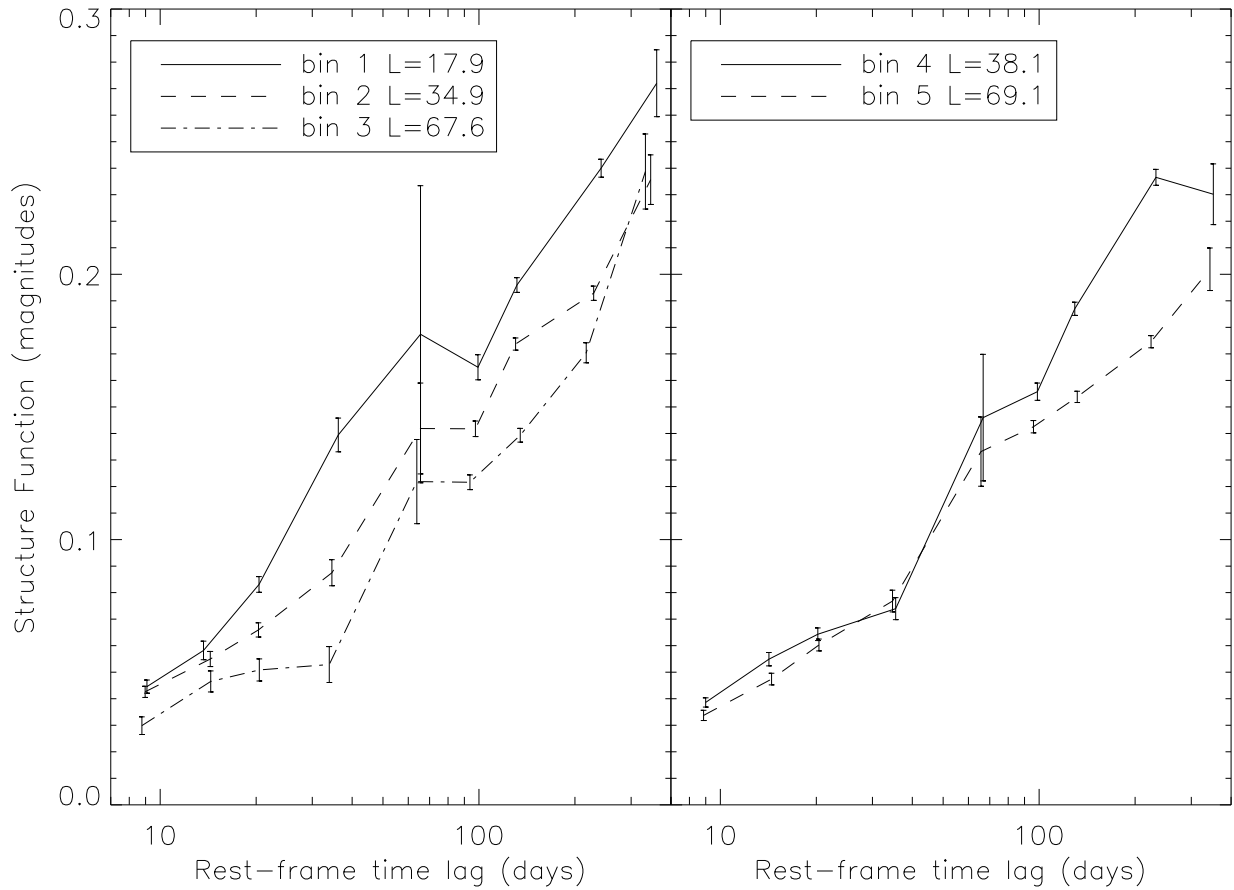}
\caption{Variability as a function of time (the Structure Function) in the $g$ band for the quasars in bins 1, 2 and 3 (left), and 4 and 5 (right).  Both panels demonstrate the well-known anticorrelation between luminosity and variability.  Luminosities listed in the legend are all $\langle\lambda{\rm L}_{\lambda} (1450)\rangle$ for than bin in units of 10$^{44}$ erg s$^{-1}$.
\label{Fig3.4}}
\end{figure}
\clearpage

\begin{figure}
\includegraphics{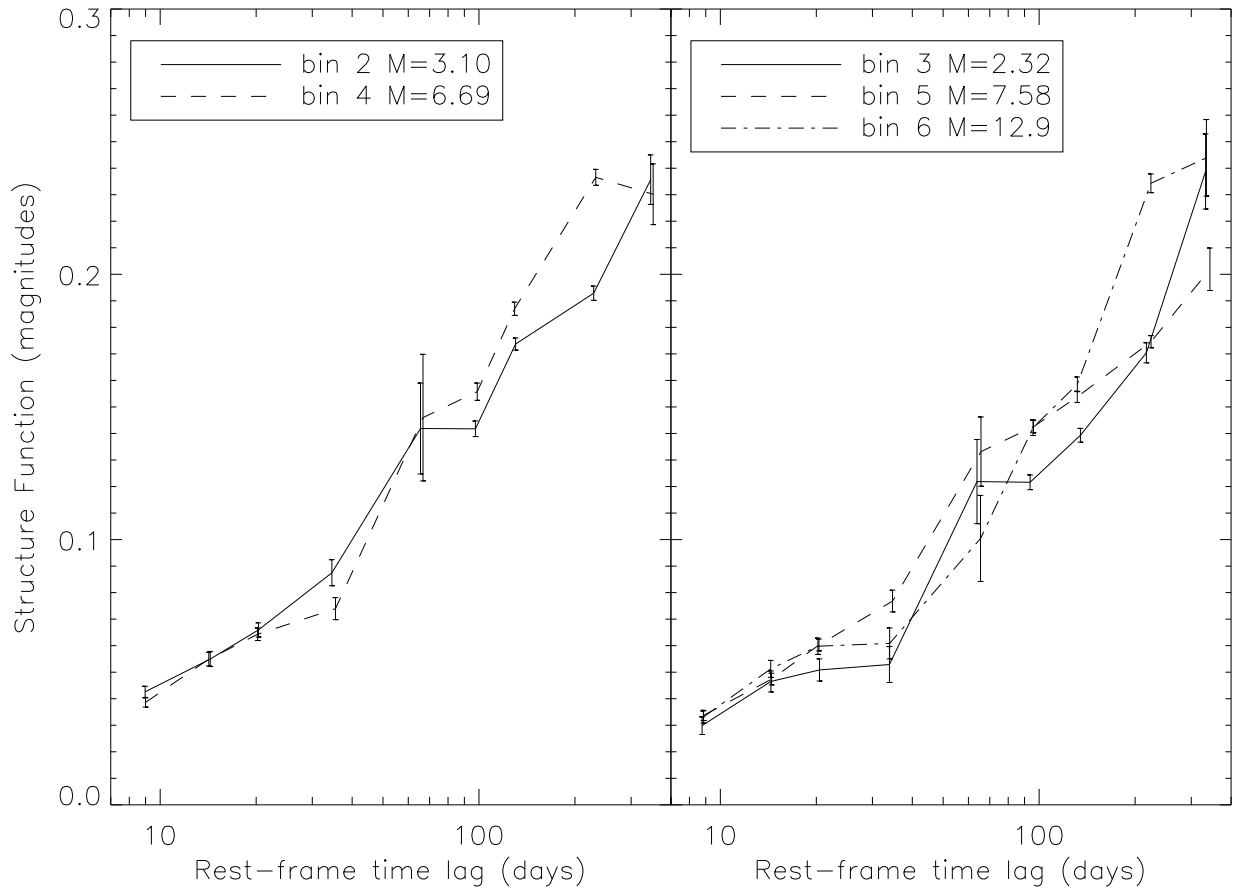}
\caption{Variability as a function of time (the Structure Function) in the $g$ band for the objects in bins 2 and 4 (left), and 3, 5 and 6 (right).  Masses listed in the legend are $\langle{\rm M}_{\rm BH}\rangle$ for that bin in units of $10^{8}$ M$_{\sun}$.
\label{Fig3.5}}
\end{figure}
\clearpage


%
%

\begin{table*}
\centering
\caption{Results of Power-Law fits to structure functions for full sample.}
\label{fullplfittab}
\begin{tabular}{crcc}
\hline
Band & $\Delta{\tau}_{0}$ (days) & $\alpha$ &  $V(\Delta{\tau} = 100)$\\
\hline

       $u$ &  5610   & 0.435 & $0.173 \pm 0.001$\\
       $g$ &  5438  & 0.479 & $0.147  \pm 0.001$\\
       $r$ &  7702   & 0.486 & $0.121  \pm 0.001$\\
       $i$ &   16490  & 0.436   & $0.108 \pm 0.001$\\
       $z$ & 33400  & 0.411 & $0.091 \pm 0.001$\\

\hline
\end{tabular}
\end{table*}

\begin{table*}
\centering
\caption{Statistics for objects in each bin in Black Hole Mass and Continuum Luminosity (as seen in Figure \ref{Fig3.2}).  Luminosities listed are all $\langle\lambda{\rm L}_{\lambda} (1450)\rangle$ in units of 10$^{44}$ erg s$^{-1}$.  Black hole masses are in units of M$_{\sun}$.
Average values listed for luminosity,  black hole mass and redshift values represent the median for that bin.}
\label{bininfotab}
\begin{tabular}{cccccccccc}
\hline
Bin & Number of & L$_{\rm low}$ & L$_{\rm high}$ & M$_{\rm low}$ & M$_{\rm high}$ &  $\langle{\rm L}\rangle$ & $\langle z\rangle$  & $\langle{\rm M}_{\rm BH}\rangle$ &  ${\rm L}/{\rm L}_{Edd}$\\
 & Objects & & & & & & & & \\
\hline

       1 & 246 &  0   & 25   & $10^{6}$                & 5 $\times 10^{8}$ & 17.9  & 1.836  & 2.55 $\times 10^{8}$ & 0.27 \\
       2 & 303 &  25 & 50   & $10^{6}$               & 5 $\times 10^{8}$ &  34.9  & 2.015  & 3.10 $\times 10^{8}$  & 0.43 \\
       3 & 190 & 50 & 100 &  $10^{6}$              & 5 $\times 10^{8}$ & 67.6  & 2.420  & 2.32 $\times 10^{8}$   & 1.12 \\
       4 &  242 & 25 & 50   & 5 $\times 10^{8}$ & $10^{9}$               & 38.1  & 1.974  & 6.69 $\times 10^{8}$   & 0.22 \\
       5 &  274 & 50 & 100 & 5 $\times 10^{8}$ & $10^{9}$               & 69.1  & 2.180  & 7.58 $\times 10^{8}$ &  0.35 \\
       6 &  191 & 50 & 100 & $10^{9}$                & $2 \times 10^{9}$ & 73.3  & 2.154  & 1.29 $\times 10^{9}$  & 0.22 \\
\hline
\end{tabular}
\end{table*}

\begin{table*}
\centering
\caption{Results of Power-Law fits to $g$-band structure functions for bins in luminosity and black hole mass.}
\label{mbhlumplfittab}
\begin{tabular}{cccc}
\hline
Bin & $\Delta{\tau}_{0}$ (days) & $\alpha$ & $V(\Delta{\tau} = 100)$\\
\hline

       1 &   3274.89 &  0.513846 & $0.167 \pm 0.002$\\
       2 &   5486.22 & 0.488012  &  $0.142 \pm 0.002$\\
       3 &   5008.55 &  0.543326 & $0.119 \pm 0.002$\\
       4 &    2811.68  &  0.561319 & $0.153 \pm 0.002$\\
      5 &  4957.32  &  0.523827  & $0.129 \pm 0.001$\\
	6 &   2940.30 & 0.578294 &  $0.141 \pm 0.002$\\

\hline
\end{tabular}
\end{table*}


\label{lastpage}


\begin{thebibliography}{99}
\bibitem[\protect\citeauthoryear{Adelman-McCarthy et al.}{2007}]{adelman07} Adelman-McCarthy, J., et 
al.\ 2007, ApJS, submitted
\bibitem[\protect\citeauthoryear{Baskin \& Laor}{2005}]{baskin05} Baskin A., Laor A., 2005, MNRAS, 356, 1029 
\bibitem[\protect\citeauthoryear{Bentz et al.}{2006}]{bentz06} 
Bentz M.~C., Peterson B.~M., Pogge R.~W., Vestergaard M., Onken C.~A., 
2006, ApJ, 644, 133 
\bibitem[\protect\citeauthoryear{Blandford \& McKee}{1982}]{blandford82} Blandford R.~D., McKee C.~F., 1982, ApJ, 
255, 419 
\bibitem[\protect\citeauthoryear{Blanton et al.}{2003}]{blanton03} Blanton, M.~R., Lin, 
 H., Lupton, R.~H., Maley, F.~M., Young, N., Zehavi, I., \& Loveday, J.\ 
  2003, AJ, 125, 2276 
\bibitem[\protect\citeauthoryear{Boller et al.}{1997}]{boller97} 
Boller T., Brandt W.~N., Fabian A.~C., Fink H.~H., 1997, MNRAS, 289, 393 
\bibitem[\protect\citeauthoryear{Cid Fernandes, Sodr{\' e}, \& Vieira da Silva}{2000}]{cid00} Cid Fernandes, R., 
  Sodr{\' e}, L., \& Vieira da Silva, L.\ 2000, ApJ, 544, 123 
\bibitem[\protect\citeauthoryear{Cristiani, Trentini, La Franca, \& Andreani}{1997}]{cristiani97} Cristiani, S., 
  Trentini, S., La Franca, F., \& Andreani, P.\ 1997, A\&A, 321, 123 
\bibitem[\protect\citeauthoryear{de Vries, Becker, \& White}{2003}]{devries03} de Vries, 
  W.~H., Becker, R.~H., \& White, R.~L.\ 2003, AJ, 126, 1217 
  \bibitem[\protect\citeauthoryear{di Clemente et al.}{1996}]{diclemente96} di Clemente A., Giallongo E., Natali G., 
Trevese D., Vagnetti F., 1996, ApJ, 463, 466 
\bibitem[\protect\citeauthoryear{Eisenstein et al.}{2001}]{eisenstein01} Eisenstein, D.~J.~et 
  al.\ 2001, AJ, 122, 2267 
\bibitem[\protect\citeauthoryear{Fukugita et al.}{1996}]{fukugita96} Fukugita, M., 
  Ichikawa, T., Gunn, J.~E., Doi, M., Shimasaku, K., \& Schneider, D.~P.\ 
  1996, AJ, 111, 1748 
\bibitem[\protect\citeauthoryear{Giveon et al.}{1999}]{giveon99} Giveon, U., Maoz, D., 
  Kaspi, S., Netzer, H., \& Smith, P.~S.\ 1999, MNRAS, 306, 637
\bibitem[\protect\citeauthoryear{Gunn et al.}{1998}]{gunn98} Gunn, J.~E.~et al.\ 1998, 
  AJ, 116, 3040 
\bibitem[\protect\citeauthoryear{Gunn et al.}{2006}]{gunn06} Gunn J.~E., et al., 2006, AJ, 131, 2332 
\bibitem[\protect\citeauthoryear{Haas}{2004}]{haas04} Haas, M.\ 2004, IAU Symposium, 222, 267 
\bibitem[\protect\citeauthoryear{Hawkins}{2002}]{hawkins02} Hawkins, M.~R.~S.\ 2002, 
  MNRAS, 329, 76 
\bibitem[\protect\citeauthoryear{Helfand et al.}{2001}]{helfand01} Helfand, D.~J., Stone, 
  R.~P.~S., Willman, B., White, R.~L., Becker, R.~H., Price, T., Gregg, 
  M.~D., \& McMahon, R.~G.\ 2001, AJ, 121, 1872 
\bibitem[\protect\citeauthoryear{Hogg, Finkbeiner, Schlegel, \& Gunn}{2001}]{hogg01} 
  Hogg, D.~W., Finkbeiner, D.~P., Schlegel, D.~J., \& Gunn, J.~E.\ 2001, AJ, 
  122, 2129 
\bibitem[\protect\citeauthoryear{Hook, McMahon, Boyle, \& Irwin}{1994}]{hook94} Hook, I.~M.,
  McMahon, R.~G., Boyle, B.~J., \& Irwin, M.~J.\ 1994, MNRAS, 268, 305
\bibitem[\protect\citeauthoryear{Hopkins et al.}{2005}]{hopkins05} Hopkins P.~F., Hernquist L., Martini P., 
Cox T.~J., Robertson B., Di Matteo T., Springel V., 2005, ApJ, 625, L71 
\bibitem[\protect\citeauthoryear{Hopkins, Richards, \& Hernquist}{2007}]{hopkins07} Hopkins P.~F., Richards G.~T., 
Hernquist L., 2007, ApJ, 654, 731 

\bibitem[\protect\citeauthoryear{Ivezi{\'c} et al.}{2004}]{ivezic04} Ivezi{\'c} {\v Z}., et al., 2004, AN, 325, 
583 
\bibitem[\protect\citeauthoryear{Kaspi et al.}{2000}]{kaspi00} Kaspi, S., Smith, P.~S., 
Netzer, H., Maoz, D., Jannuzi, B.~T., \& Giveon, U.\ 2000, ApJ, 533, 631 
\bibitem[\protect\citeauthoryear{Kaspi et al.}{2005}]{kaspi05} 
Kaspi S., Maoz D., Netzer H., Peterson B.~M., Vestergaard M., Jannuzi 
B.~T., 2005, ApJ, 629, 61 
\bibitem[\protect\citeauthoryear{Kelly \& Bechtold}{2007}]{kelly07} Kelly B.~C., Bechtold J., 2007, ApJS, 168, 1 

\bibitem[\protect\citeauthoryear{Kinman}{1968}]{kinman68} Kinman 
T.~D., 1968, Sci, 162, 1081 
\bibitem[\protect\citeauthoryear{Kollmeier et al.}{2006}]{kollmeier06} Kollmeier J.~A., et al., 2006, ApJ, 648, 128
\bibitem[\protect\citeauthoryear{Koo, Kron, \& Cudworth}{1986}]{koo86} Koo, D.~C., 
  Kron, R.~G., \& Cudworth, K.~M.\ 1986, PASP, 98, 285 
\bibitem[\protect\citeauthoryear{Krolik}{1998}]{krolikbook} Krolik, J.~H.\ 1998, Active 
Galactic Nuclei: From the Central Black Hole to the Galactic Environment, 
by J.H.~Krolik.~Princeton: Princeton University Press, 1998
\bibitem[\protect\citeauthoryear{Lundgren et al.}{2007}]{lundgren07} Lundgren B.~F., Wilhite B.~C., Brunner 
R.~J., Hall P.~B., Schneider D.~P., York D.~G., Vanden Berk D.~E., Brinkmann J., 2007, ApJ, 656, 73 
\bibitem[\protect\citeauthoryear{Lupton, Gunn, \& Szalay}{1999}]{lupton99} Lupton R.~H., Gunn J.~E., Szalay A.~S., 
1999, AJ, 118, 1406 
\bibitem[\protect\citeauthoryear{Lupton et~al.}{2001}]{lupton01}
  Lupton, R., Gunn, J.~E., Ivezi\'c, Z., Knapp, G.~R., Kent, S.,
  \& Yasuda, N. 2001, in ASP Conf. Ser. 238, Astronomical Data Analysis
  Software and Systems X, ed. F.~R. Harnden, Jr., F.~A. Primini,
  and H.~E. Payne (San Francisco: Astr. Soc. Pac.), p. 269,
  astro-ph[0101420]
\bibitem[\protect\citeauthoryear{Martini \& Schneider}{2003}]{martini03} Martini P., Schneider D.~P., 2003, 
ApJ, 597, L109 
\bibitem[\protect\citeauthoryear{Oke \& Gunn}{1983}]{oke83} Oke, J.~B.~\& Gunn, J.~E.\ 
  1983, ApJ, 266, 713 
\bibitem[\protect\citeauthoryear{Pereyra et al.}{2006}]{pereyra06} Pereyra, N.~A., Vanden 
Berk, D.~E., Turnshek, D.~A., Hillier, D.~J., Wilhite, B.~C., Kron, R.~G., Schneider, D.~P., \& Brinkmann, J.\ 2006, ApJ, 642, 87 
\bibitem[\protect\citeauthoryear{Peterson}{1993}]{peterson93} Peterson, B.~M.\ 1993, PASP, 
105, 247 
\bibitem[\protect\citeauthoryear{Peterson et al.}{2004}]{peterson04} Peterson B.~M., et al., 2004, ApJ, 613, 
682 
\bibitem[\protect\citeauthoryear{Pier et al.}{2003}]{pier03} Pier, J.~R., et~al.\ 2003, AJ, 125, 1559
\bibitem[\protect\citeauthoryear{Pogge \& Peterson}{1992}]{pogge92} Pogge, R.~W.~\& Peterson, B.~M.\ 1992, AJ, 103, 1084
\bibitem[\protect\citeauthoryear{Rees}{1984}]{rees84} Rees M.~J., 1984, ARA\&A, 22, 471 
\bibitem[\protect\citeauthoryear{Reichard et al.}{2003}]{reichard03} Reichard, T.~A., et 
al.\ 2003, AJ, 126, 2594 
\bibitem[\protect\citeauthoryear{Rengstorf et al.}{2004}]{rengstorf04} Rengstorf A.~W., et al., 2004, ApJ, 617, 
184 
\bibitem[\protect\citeauthoryear{Rengstorf, Brunner, \& Wilhite}{2006}]{rengstorf06} Rengstorf A.~W., Brunner R.~J., 
Wilhite B.~C., 2006, AJ, 131, 1923 

\bibitem[\protect\citeauthoryear{Richards et al.}{2002}]{richards02} Richards, G.~T.~et 
al.\ 2002, AJ, 123, 2945 
\bibitem[\protect\citeauthoryear{Richards et al.}{2006}]{richards06} Richards G.~T., et al., 2006, ApJS, 166, 470 

\bibitem[\protect\citeauthoryear{Schlegel, Finkbeiner, \& Davis}{1998}]{schlegel98} 
  Schlegel, D.~J., Finkbeiner, D.~P., \& Davis, M.\ 1998, ApJ, 500, 525 
\bibitem[\protect\citeauthoryear{Shakura \& Sunyaev}{1973}]{shakura73} Shakura, N.~I. \& Sunyaev, R.~A. 1973, A\&A, 24, 337 
\bibitem[\protect\citeauthoryear{Shang et al.}{2007}]{shang07} 
Shang Z., Wills B.~J., Wills D., Brotherton M.~S., 2007, AJ, 134, 294 
\bibitem[\protect\citeauthoryear{Smith et al.}{2002}]{smith02} Smith, J.~A.~et al.\ 
  2002, AJ, 123, 2121 
\bibitem[\protect\citeauthoryear{Spergel et al.}{2006}]{spergel07} Spergel D.~N., et al., 2006, astro, 
arXiv:astro-ph/0603449 
\bibitem[\protect\citeauthoryear{Stoughton et al.}{2002}]{stoughton02} Stoughton, C.~et al.\ 
  2002, AJ, 123, 485 
\bibitem[\protect\citeauthoryear{Strauss et al.}{2002}]{strauss02} Strauss, M.~A.~et al.\ 
  2002, AJ, 124, 1810 
\bibitem[\protect\citeauthoryear{Terrell}{1967}]{terrell67} 
Terrell J., 1967, ApJ, 147, 827 
\bibitem[\protect\citeauthoryear{Tr{\` e}vese, Kron, \& Bunone}{2001}]{trevese01}
  Tr{\`e}vese, D., Kron, R.~G., \& Bunone, A.\ 2001, ApJ, 551, 103
  \bibitem[\protect\citeauthoryear{Tucker et al.}{2006}]{tucker06} Tucker D.~L., et al., 2006, AN, 327, 821 
\bibitem[\protect\citeauthoryear{Uomoto, Wills, \& Wills}{1976}]{uomoto76} Uomoto, A.~K., 
  Wills, B.~J., \& Wills, D.\ 1976, AJ, 81, 905 
\bibitem[\protect\citeauthoryear{Vanden Berk et al.}{2001}]{vandenberk01} Vanden Berk, 
  D.~E.~et al.\ 2001, AJ, 122, 549 
\bibitem[\protect\citeauthoryear{Vanden Berk et al.}{2004}]{vandenberk04} Vanden Berk, D.~E., 
et al.\ 2004, ApJ, 601, 692 
\bibitem[\protect\citeauthoryear{Vanden Berk et al.}{2005}]{vandenberk05} Vanden Berk, D.~E., 
et al.\ 2005, AJ, 129, 2047 
\bibitem[\protect\citeauthoryear{Vestergaard}{2002}]{vestergaard02} Vestergaard, M.\ 2002, 
ApJ, 571, 733 

\bibitem[\protect\citeauthoryear{Vestergaard \& Peterson}{2006}]{vestergaard06} Vestergaard, 
M., \& Peterson, B.~M.\ 2006, ApJ, 641, 689 
\bibitem[\protect\citeauthoryear{Wilhite et al.}{2005}]{wilhite05} Wilhite, B.~C., Vanden 
Berk, D.~E., Kron, R.~G., Schneider, D.~P., Pereyra, N., Brunner, R.~J., 
Richards, G.~T., \& Brinkmann, J.~V.\ 2005, ApJ, 633, 638 
\bibitem[\protect\citeauthoryear{Wilhite et al.}{2006}]{wilhite06} Wilhite B.~C., Vanden Berk D.~E., Brunner 
R.~J., Brinkmann J.~V., 2006, ApJ, 641, 78 
\bibitem[\protect\citeauthoryear{Wilkes}{1984}]{wilkes84} Wilkes, B.~J.\ 1984, MNRAS, 
207, 73 
\bibitem[\protect\citeauthoryear{Wold, Brotherton, \& Shang}{2007}]{wold07} Wold M., Brotherton M.~S., Shang Z., 
2007, MNRAS, 1496 
\bibitem[\protect\citeauthoryear{York et al.}{2000}]{york00} York, D.~G.~et al.\ 2000, 
  AJ, 120, 1579

\end{thebibliography}
\end{document}